\documentclass[conference]{IEEEtran}
\IEEEoverridecommandlockouts
\usepackage{cite}
\usepackage{amsmath,amssymb,amsfonts}
\usepackage{algorithmic}
\usepackage{textcomp}
\usepackage{xcolor}
\usepackage[pdftex]{graphicx}
\usepackage{epstopdf}

\usepackage{amsmath,amssymb,enumerate} 
\usepackage{bm} 
\usepackage{indentfirst} 
\usepackage{empheq} 
\usepackage{here} 
\newcommand{\dd}{\mathrm{d}}

\newcommand{\ii}{\mathrm{i}}

\def\BibTeX{{\rm B\kern-.05em{\sc i\kern-.025em b}\kern-.08em
    T\kern-.1667em\lower.7ex\hbox{E}\kern-.125emX}}
\begin{document}

\title{Polarization Model of Online Social Networks \\
Based on the Concept of \\
Spontaneous Symmetry Breaking\\
}

\author{\IEEEauthorblockN{Masaki Aida}
    \IEEEauthorblockA{
        \textit{Tokyo Metropolitan University} \\
Tokyo 191--0065, Japan \\
aida@tmu.ac.jp}
\and
\IEEEauthorblockN{Ayako Hashizume}
    \IEEEauthorblockA{
        \textit{Hosei University} \\
Tokyo 194--0298, Japan \\
hashiaya@hosei.ac.jp}
\and
\IEEEauthorblockN{Chisa Takano}
    \IEEEauthorblockA{
        \textit{Hiroshima City University} \\
Hiroshima 731-3194, Japan \\
takano@hiroshima-cu.ac.jp}
\and
\IEEEauthorblockN{Masayuki Murata}
    \IEEEauthorblockA{
        \textit{Osaka University} \\
Osaka, 565-0871 Japan \\
murata@ist.osaka-u.ac.jp}
}

\maketitle

\begin{abstract}
The spread of information networks has not only made it easier for people to access a variety of information sources but also greatly enhanced the ability of individuals to disseminate information.
Unfortunately, however, the problem of slander in online social networks shows that the evolving information network environment does not necessarily support mutual understanding in society. 
Since information with particular bias is distributed only to those communities that prefer it, the division of society into various opposing groups is strengthened. 
This phenomenon is called polarization.
It is necessary to understand the mechanism of polarization to establish technologies that can counter polarization.
This paper introduces a fundamental model for understanding polarization that is based on the concept of spontaneous symmetry breaking; our starting point is the oscillation model that describes user dynamics in online social networks. 
\end{abstract}


\section{Introduction}
\label{sec:firstly}
In recent decades, although the development of digital technologies and social media has activated information exchange among users and has supported a variety of social activities, it has, unfortunately, offset its benefits in terms of social activity with very negative problems.  
For example, social polarization and the incitement of extremist violence are likely empowered by online social networks (OSNs). 
Of particular import is the echo-chamber effect, whereby users in a polarized community develop extremely biased opinions. 

Social polarization and the echo-chamber effect have been investigated using real data of information networks; 
\cite{Flaxman} analyzed Web browsing log data gathered from SNS and search engines, 
\cite{Bessi} analyzed Facebook data, while \cite{Bessi2} examined both Facebook and YouTube data. 
By analyzing Twitter data, \cite{Garimella} identified topics that were likely to activate the discussion, 
\cite{Williams} used the topic of climate change to analyze the echo-chamber effect, while \cite{Baumann} proposed a model of social polarization based on an ordinary differential equation. 

These studies phenomenologically consider the process of social polarization and the occurrence of the echo-chamber effect, based on actual data.
They did not theoretically consider what kind of user dynamics will occur on OSN after social polarization occurs.
To establish a complete theoretical model of the echo chamber phenomenon, it is indispensable to understand the user dynamics that result from social polarization.

Spontaneous symmetry breaking is a theoretical model used to describe the spontaneous magnetization of a ferromagnet.
The atoms of ferromagnetic material are themselves small magnets and no overall magnetism is exhibited if each atom is oriented in a random direction. 
However, if the directions of the atoms become aligned for some reason, the material acts as a single large magnet. 
Social polarization exhibits strong parallels with spontaneous symmetry breaking because many users become aligned with a particular opinion \cite{Galam}.

Spontaneous symmetry breaking is essentially a theoretical model in the framework of quantum theory, which makes it is difficult to derive useful engineering methods to tackle the echo-chamber effect. 
For example, spontaneous symmetry breaking in quantum theory is known to generate new dynamics (called the Nambu-Goldstone mode) that did not exist before symmetry breaking.
However, we, unfortunately, do not understand what the Nambu-Goldstone mode corresponds to in real OSNs. 
This means that trying to draw a superficial analogy with symmetry breaking is problematic. 

This paper draws on quantum theory and spontaneous symmetry breaking to develop a theoretical model of user dynamics via OSNs. We tackle the problem of estimating what new user dynamics are likely to be triggered by social polarization. 

\section{The Oscillation Model for User Dynamics}
\begin{figure*}
\begin{center}
\begin{tabular}{ccccc}
\includegraphics[width=0.255\linewidth]{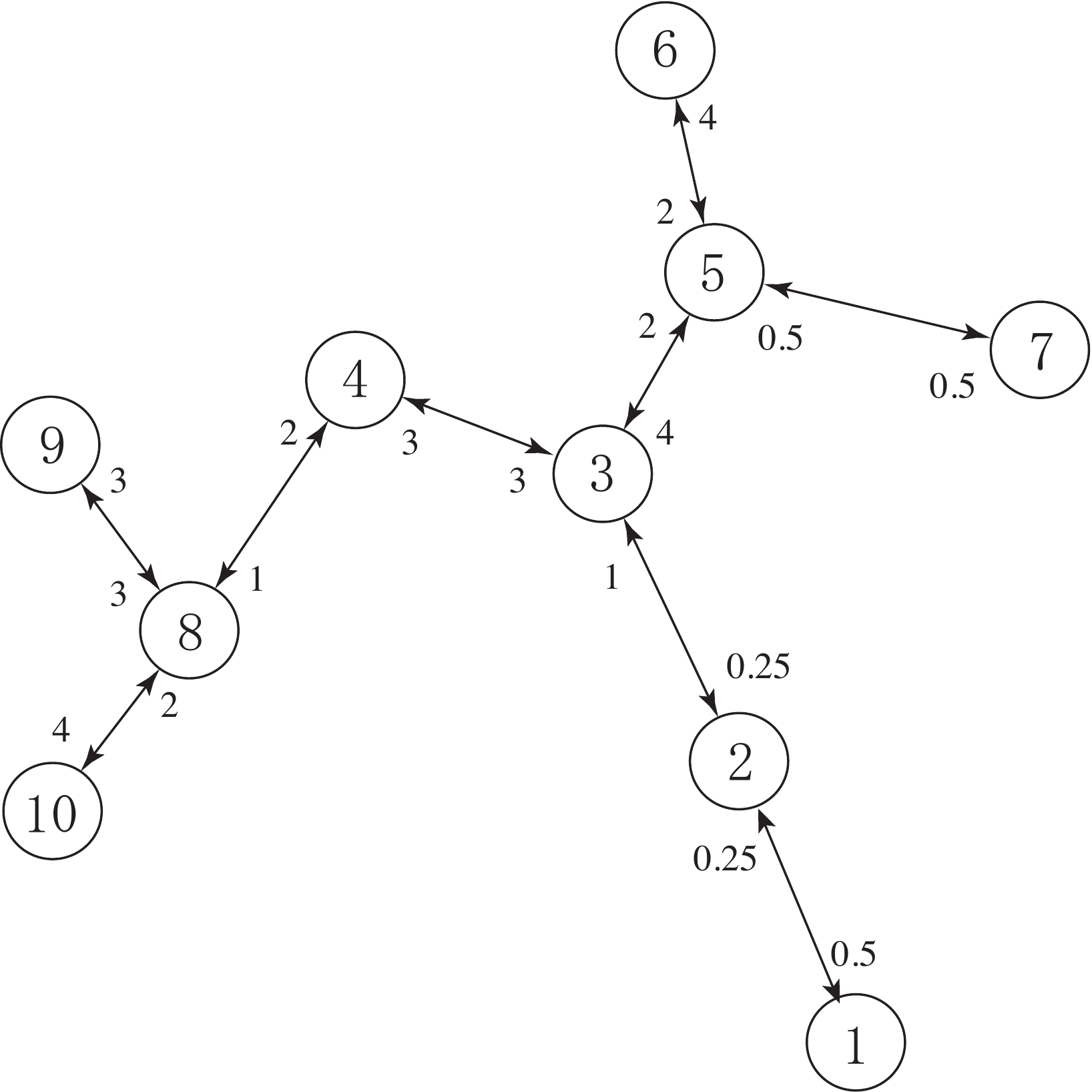} & \quad & 
\includegraphics[width=0.31\linewidth]{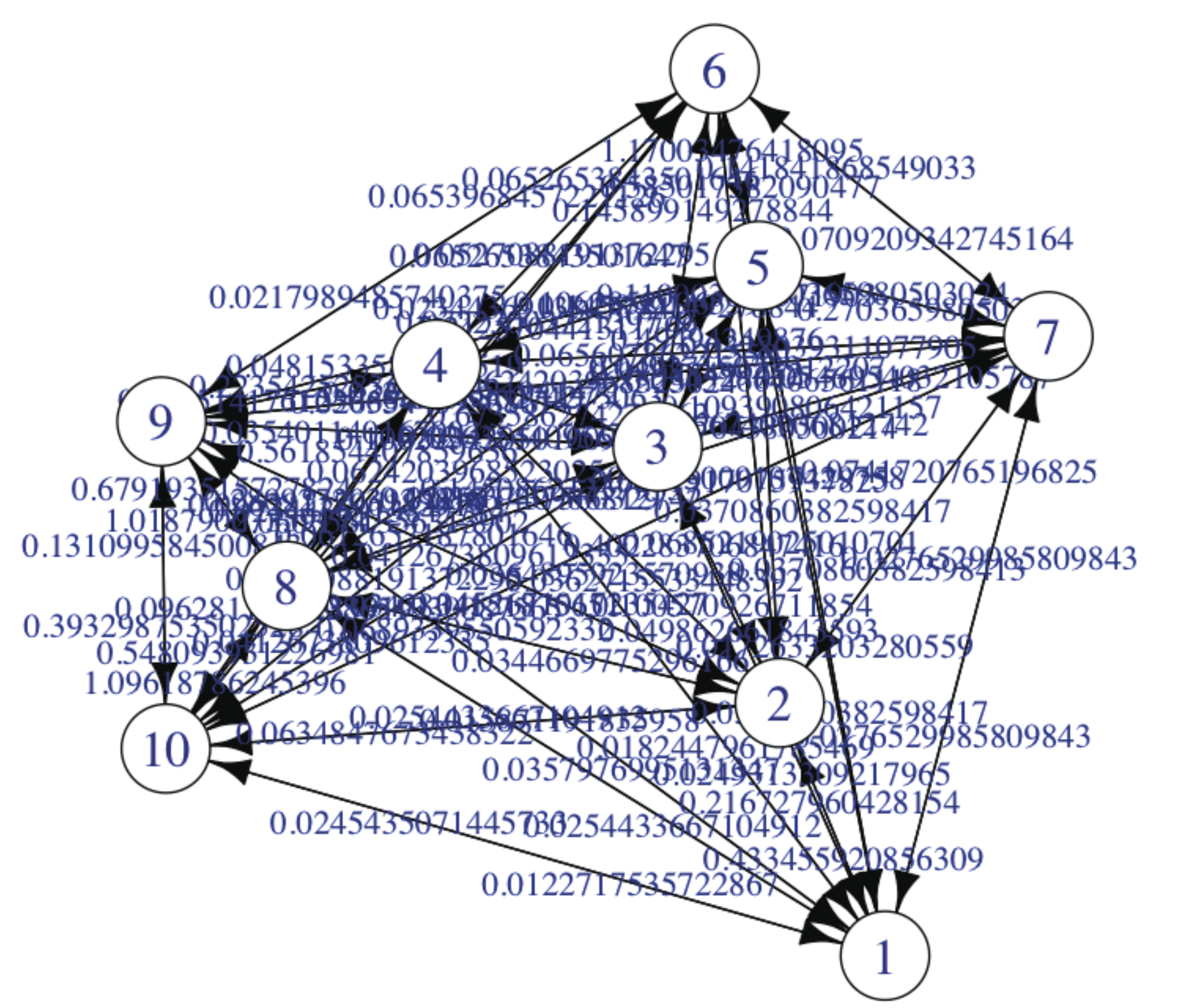} & \quad & 
\includegraphics[width=0.255\linewidth]{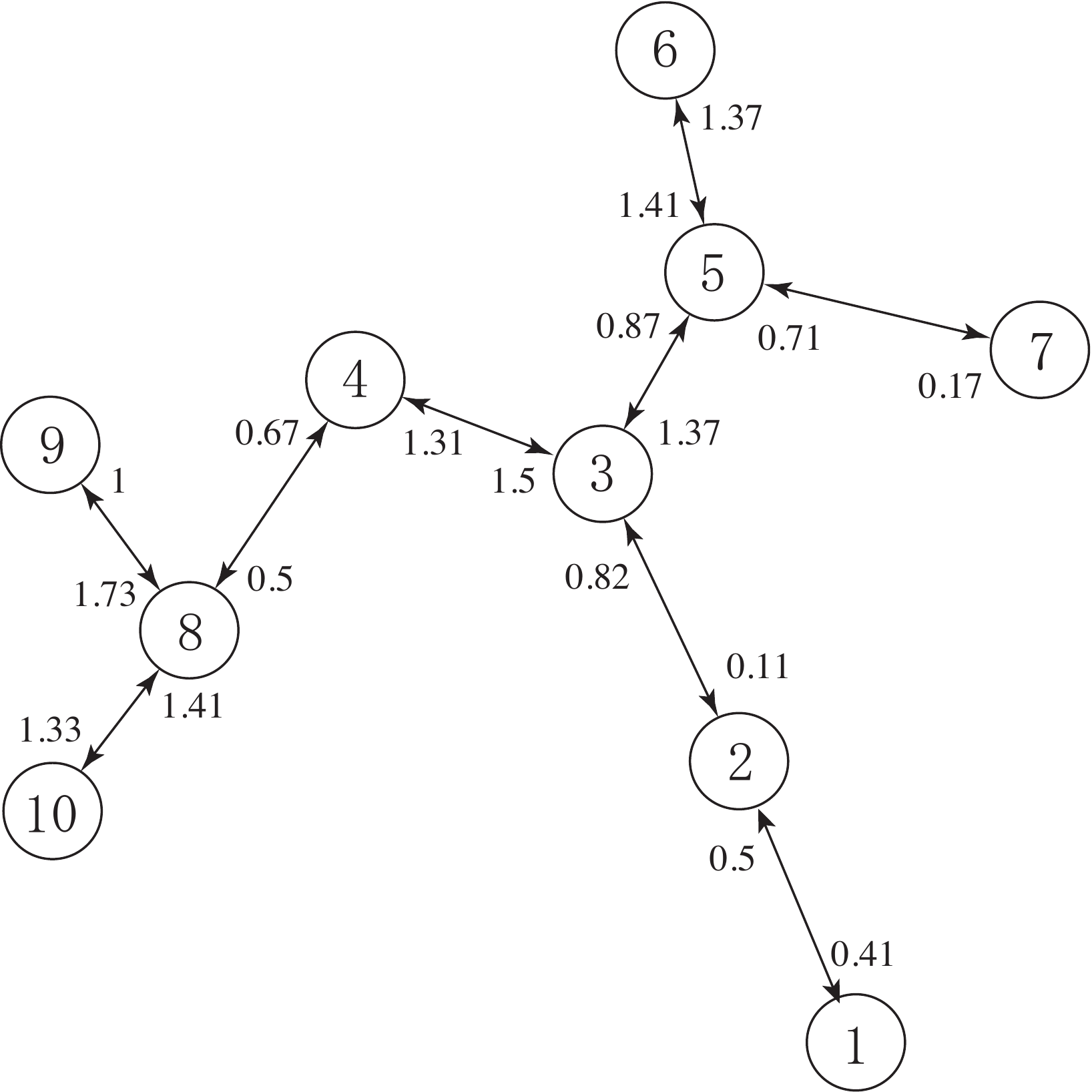}\\
Laplacian matrix $\bm{\mathcal{L}}$ && square root matrix $\sqrt{\bm{\mathcal{L}}}$ && semi-normalized Laplacian matrix $\bm{\mathcal{H}}$
\end{tabular}
\caption{An example of network structures represented by the Laplacian matrix $\bm{\mathcal{L}}$, 
the square root matrix $\sqrt{\bm{\mathcal{L}}}$, and the Hamiltonian $\bm{\mathcal{\Hat{H}}}$}
\label{fig:1}
\end{center}
\end{figure*}
The oscillation model was proposed to describe user dynamics in OSNs ~\cite{aida-book2020,aida2018}.
A key feature of this model is that it is not intended to provide a complete description of the complex thoughts and behaviors of users, 
rather, a minimal model of user interaction is assumed that is as simple and universal as possible. 
The wave equation in OSNs derived therefrom can describe the characteristics of user dynamics shared by many users.
Although the oscillation model is simple, it has been confirmed that it has the following advantages.
\begin{enumerate}[$\bullet$]
\item The oscillation energy for each node calculated from the oscillation model can provide not only a common framework 
yielding the conventional indices of node centrality (degree centrality and betweenness centrality) \cite{wasserman,carrington,mislove} 
but also can an extend the concept of node centrality to encompass more complex configurations of OSNs 
\cite{takano2018,takano2016,aida2016}. 
\item It is possible to describe the phenomenon of an explosion in the activity of user dynamics as exemplified by flaming in OSNs~\cite{aida2018,aida2017}.
\item The oscillation model predicts that low-frequency oscillation modes will become dominant when the OSNs are activated. 
The prediction was confirmed by an analysis of actual data on Google Trends and ``2 channel'' (the largest Japanese electric bulletin board system) \cite{nagatani_infocom_2019,nagatani_bigdata_2019}. 
\item The oscillation model not only provides the solution that describes the user dynamics but also elucidates the causal relationship of the effect of OSN structure on user dynamics~\cite{aida_netscix_2020}.
The fundamental equation that describes the user dynamics and the causal relationship is similar in form to the Dirac equation, which is well known in relativistic quantum mechanics.
\end{enumerate}

The fundamental equations of user dynamics can be summarized as follows.
First, for nodes $i,\,j \in V$ of directed graph $G (V, E)$ representing the structure of an OSN with $n$ users, if the weight of directed link $(i \rightarrow j) \in E$ is given as $ w_ {ij} $, the adjacent matrix $\bm{\mathcal{A}} = [\mathcal{A}_{ij}]_{1\le i,j \le n}$ is defined as 
\begin{align}
\mathcal{A}_{ij} := \left\{
\begin{array}{cl}
w_{ij},&  \quad (i\rightarrow j) \in E,\\
0,& \quad (i\rightarrow j) \not\in E.
\end{array}
\right. 
\end{align}
Also, given nodal (weighted) out-degree $d_i := \sum_{j\in \partial i} w_{ij}$, 
the degree matrix is defined as 
\begin{align}
\bm{\mathcal{D}} := \mathrm{diag}(d_1,\,\dots\,d_n).
\end{align}
Here, $\partial i$ denotes the set of adjacent nodes of out-links from node $i$. 
Next, the Laplacian matrix of the directed graph representing the structure of the OSN is defined 
by 
\begin{align}
\bm{\mathcal{L}} := \bm{\mathcal{D}} - \bm{\mathcal{A}}.
\end{align}

Let the state vector of users at time $t$ be 
\[
\bm{x}(t) := {}^t\!(x_1(t),\,\dots,\,x_n(t)), 
\]
where 
$x_i(t)$ $(i=1,\,\dots,\,n)$ denotes the user state of node $i$ at time $t$. 
Then, the wave equation for the OSN is written as 
\begin{align}
\frac{\dd^2}{\dd t^2} \, \bm{x}(t) = - \bm{\mathcal{L}} \, \bm{x}(t). 
\label{eq:wave-eq}
\end{align}
Here, in addition to simply finding the solution $\bm{x}(t)$ of the wave equation (\ref{eq:wave-eq}), 
it is desirable to be able to describe what kind of OSN structure impacts user dynamics. 
In other words, we want to describe the causal relationship between OSN structure and user dynamics. 
To achieve this, we need to develop that is a first-order differential equation with respect to time 
(hereinafter referred to as the fundamental equation) ~\cite{aida-book2020,aida2018}.
By using the semi-definite matrix $\sqrt{\bm{\mathcal{L}}}$, which is the square root of $\bm{\mathcal {L}}$, 
we introduce the following fundamental equation, 
\begin{align}
\pm\ii \,\frac{\dd}{\dd t}\, \bm{x}^\pm(t) &= \sqrt{\bm{\mathcal{L}}} \, \bm{x}^\pm(t) \quad\text{(double-sign corresponds)},
\label{eq:fundamental-0}
\end{align}
where $\sqrt{\bm{\mathcal{L}}}$ is uniquely determined for $\bm{\mathcal {L}}$. 
The two fundamental equations (\ref{eq:fundamental-0}) can, by using a $2n$-dimensional state vector, be expressed as a single expression 
of 
\begin{align}
\ii \,\frac{\dd}{\dd t}\, \bm{\Hat{x}}(t) = \left( \sqrt{\bm{\mathcal{L}}} \otimes 
\begin{bmatrix}
+1 & 0\\
0 & -1
\end{bmatrix}
 \right) \bm{\Hat{x}}(t),
\label{eq:fundamental-1}
\end{align}
where for $\bm{x}^\pm(t) = {}^t\!(x_1^\pm(t),\,\dots,\,x_n^\pm(t))$ (double-sign corresponds), 
the $2n$-dimensional state vector $\bm{\Hat{x}}(t)$ is defined as 
\[
\bm{\Hat{x}}(t) := \bm{x}^+(t) \otimes 
\begin{pmatrix}
1\\
0
\end{pmatrix}
+ \bm{x}^-(t) \otimes 
\begin{pmatrix}
0\\
1
\end{pmatrix}. 
\]
Also, $\otimes$ denotes the Kronecker product\cite{brewer}. 

\begin{figure*}
\begin{center}
\includegraphics[width=0.60\linewidth]{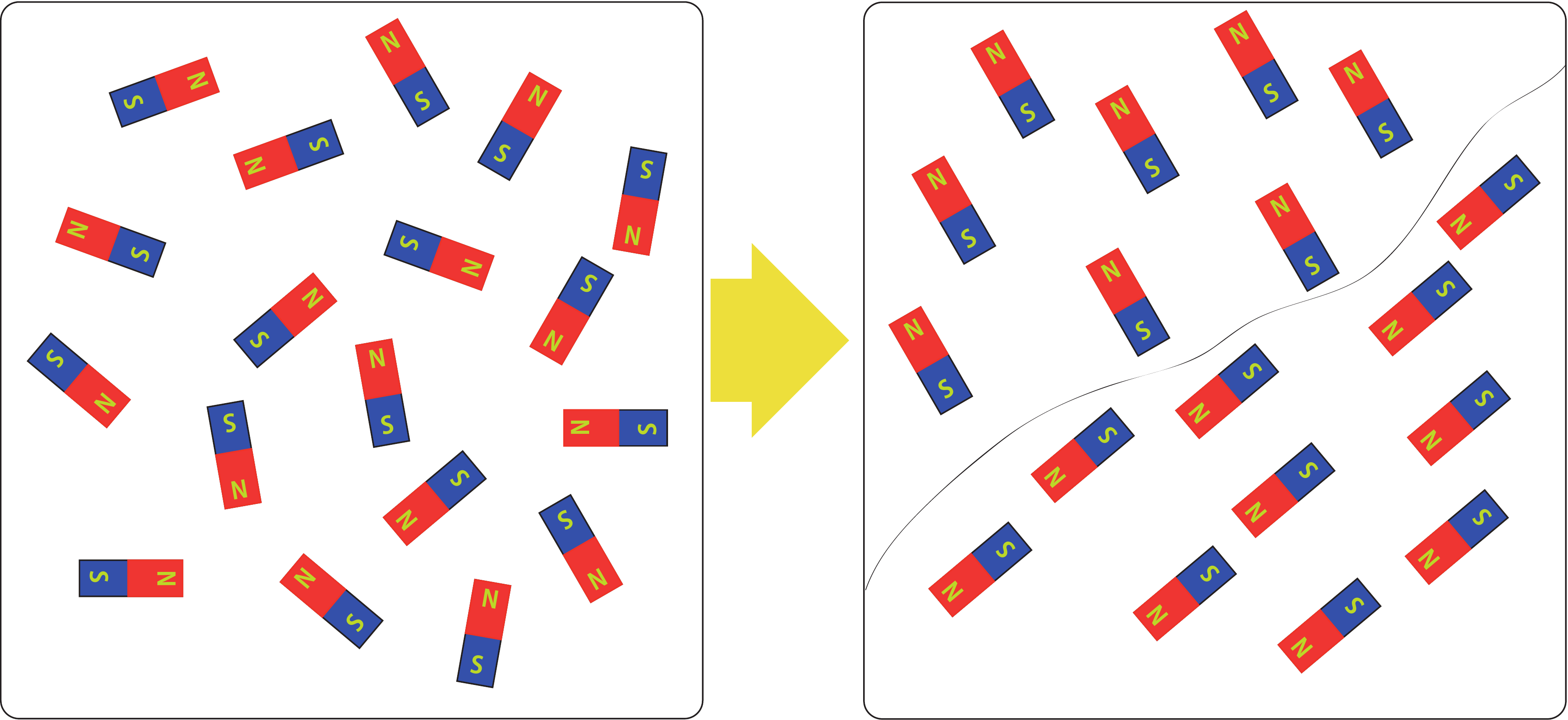}
\caption{Image of spontaneous magnetization of ferromagnetic material}
\label{fig:mag}
\end{center}
\end{figure*}
The fundamental equation~(\ref{eq:fundamental-1}) is considered to be more theoretically useful than the original wave equation (\ref{eq:wave-eq}) 
because it can describe the causal relationship of how OSN structure influences user dynamics. 
However, there is the following technical issue.
Even if the Laplacian matrix $\bm{\mathcal {L}}$ is sparse, its square root matrix $\sqrt{\bm{\mathcal{L}}}$ is generally a complete graph (see Fig.~\ref{fig:1}).
In ordinary large-scale social networks, it is implausible to assume a situation in which all users are connected. 
The matrix that appears in the fundamental equation must completely reflect the OSN link structure (whether there is a link between OSN nodes).

According to the literature~\cite{aida-book2020, aida_netscix_2020}, we can derive fundamental equations 
to solve the above problem.
The method is as follows.

Let us introduce a new matrix $\bm{\mathcal{H}}$ as follows. 
\begin{align}
\bm{\mathcal{H}} := \sqrt{\bm{\mathcal{D}}^{-1}} \, \bm{\mathcal{L}} = \sqrt{\bm{\mathcal{D}}} - \sqrt{\bm{\mathcal{D}}^{-1}} \, \bm{\mathcal{A}}, 
\label{eq:def_H}
\end{align}
where $\sqrt{\bm{\mathcal{D}}} := \text{diag}(\sqrt{d_1},\,\dots,\,\sqrt{d_n})$. 
As is well-known, the normalized Laplacian matrix is defined as
\[
\bm{\mathcal{N}} := \sqrt{\bm{\mathcal{D}}^{-1}}\, \bm{\mathcal{L}} \,  \sqrt{\bm{\mathcal{D}}^{-1}} = \bm{I} -   \sqrt{\bm{\mathcal{D}}^{-1}} \, \bm{\mathcal{A}}\, \sqrt{\bm{\mathcal{D}}^{-1}}; 
\]
so we call $\bm{\mathcal{H}}$ the semi-normalized Laplacian matrix. 
Here, $\bm{I}$ is the $n\times n$ unit matrix. 

Using the semi-normalized Laplacian matrix $\bm{\mathcal{H}}$, a new fundamental equation for user dynamics can be written as follows.
\begin{align}
\ii \, \frac{\dd}{\dd t}\, \bm{\Hat{x}}(t) &= \bm{\mathcal{\Hat{H}}}\, \bm{\Hat{x}}(t), 
\label{fundamenrtal-eq_std}
\end{align}
where $\bm{\mathcal{\Hat{H}}}$ is a $2n\times 2n$ matrix and is defined by using the Kronecker product 
as follows. 
\begin{align}
 \bm{\mathcal{\Hat{H}}} &:= 
\sqrt{\bm{\mathcal{D}}} \otimes
\begin{bmatrix}
+1 & 0\\
0 & -1
\end{bmatrix}
- \left(\sqrt{\bm{\mathcal{D}}^{-1}} \, \bm{\mathcal{A}}\right) \otimes \frac{1}{2}
\begin{bmatrix}
+1 & +1\\
-1 & -1
\end{bmatrix}. 
\label{eq:hat_H}
\end{align}
Since matrix $\bm{\mathcal{\Hat{H}}}$ governs temporal evolution 
in the fundamental equation~(\ref{fundamenrtal-eq_std}), 
we call $\bm{\mathcal{\Hat{H}}}$ the Hamiltonian. 
Also, for $n$-dimensional vector $\bm{x}(t) := {}^t\!(x_1(t),\,\dots,\,x_n(t))$, 
$\bm{\Hat{x}}(t)$ is a $2n$-dimensional vector, and their relationship is expressed as 
\begin{align}
\bm{x}(t) = (\bm{I}\otimes (1,1))\,\bm{\Hat{x}(t)}. 
\label{eq:x}
\end{align}

The graph structure (whether there is a link between nodes) represented by the Hamiltonian $\bm{\mathcal{\Hat {H}}}$ is exactly the same as the Laplacian matrix $\bm{\mathcal{L}}$ of the OSN (see Fig.~\ref {fig:1}).
In addition, the solution $\bm{x}(t)$ obtained from the fundamental equation~(\ref{fundamenrtal-eq_std}) and the relation (\ref{eq:x}) is also a solution of the original wave equation~(\ref{eq:wave-eq}).

\section{Polarization and Spontaneous Symmetry Breaking}
\begin{figure*}
\begin{center}
\begin{tabular}{ccc}
\includegraphics[width=0.35\linewidth]{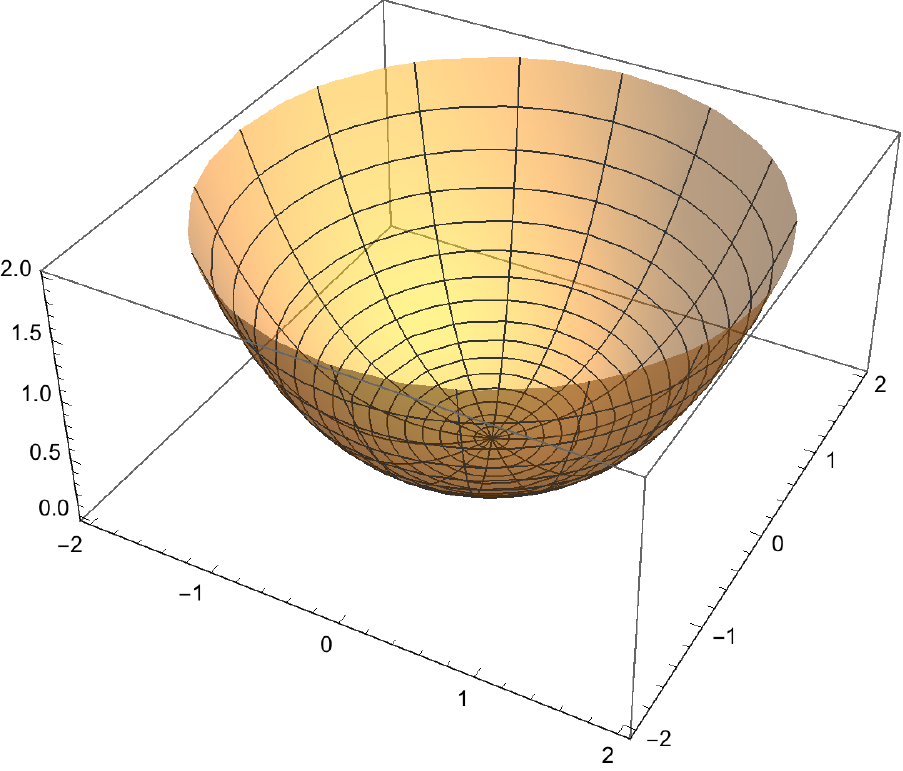}& \vspace{2mm}
&\includegraphics[width=0.35\linewidth]{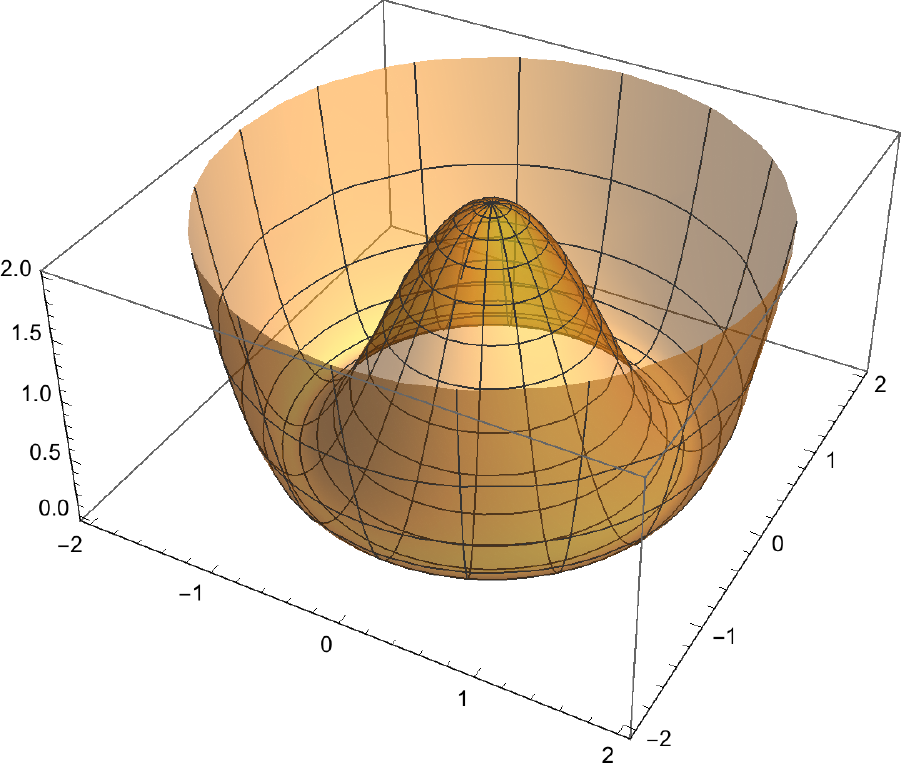}
\vspace{-2mm}
\\
{\small (a) Before symmetry breaking} & & {\small (b) After symmetry breaking}
\end{tabular}
\caption{The shapes of the potential function before and after spontaneous symmetry breaking }
\label{fig:SSB}
\end{center}
\vspace{3mm}
\end{figure*}
Spontaneous symmetry breaking refers to a situation in which an asymmetric state emerges that is stable as a result of the symmetry originally possessed by the system breaking spontaneously.
A typical example is a spontaneous magnetization that occurs in a specific direction when the spins of magnetic atoms come to align in that direction. 
Figure~\ref{fig:mag} illustrates two domains being created by the spontaneous magnetization of a ferromagnetic material. 
Since the alignment of atoms occurs by local interaction and grows up to the domain boundaries, the entire ferromagnetic material is not magnetized in completely the same direction.
A metaphor was introduced that linked polarization by structural changes in OSNs to spontaneous symmetry breaking~\cite{Galam}.
However, since spontaneous symmetry breaking is a concept in the framework of quantum theory, 
it has been difficult to apply it mathematically in models of social systems.
As described later, the oscillation model is naturally linked to the framework of quantum theory as the fundamental equation representing user dynamics, so there is the possibility that we can naturally and usefully incorporate the concept of spontaneous symmetry breaking into the oscillation model. 

We briefly summarize a method for introducing the concept of spontaneous symmetry breaking into quantum theory as follows.
Assuming that spontaneous symmetry breaking is introduced for a specific oscillation mode $\mu$ appearing in the fundamental equation (\ref{fundamenrtal-eq_std}) of the oscillation model, let the two solutions appearing in the oscillation mode be $\psi_\mu^+(t)$ and $\psi_\mu^-(t)$; we define $|\psi_\mu|^2 := |\psi_\mu^+|^ 2 + |\psi_\mu ^-|^2$. 
Then we introduce potential function $V(|\psi_\mu|^2)$ as
\begin{align}
V(|\psi_\mu|^2) = a \, |\psi_\mu|^2 + b \, |\psi_\mu|^4 + \mathrm{const.}
\label{eq:potential}
\end{align}
The shape of the potential function is formed by terms of even powers of $|\psi_\mu|$ so as to provide a stable ground state, and higher-order terms are negligible in the region where $|\psi_\mu|$ is small. 
The constant term is introduced to adjust the value of the lowest point (ground state) of the potential function to $0$.
Here, for a given constant $b> 0$, if $a \ge 0$, the shape of the potential function is as shown in Fig.~\ref{fig:SSB} (a).
If $ a <0 $, the shape is like the punt of a wine bottle, see Fig.~\ref{fig:SSB} (b); this is called the Mexican-hat potential.

If $a \ge 0$, $|\psi_\mu| = 0$ is stable, but if $a <0$, it is stable at $|\psi_\mu| = \sqrt{|a|/2b}$. 
Then one of the infinitely many stable states is spontaneously selected; this constitutes an example of spontaneous symmetry breaking.
When $|\psi_\mu| \not = 0$ is in a stable state, its oscillation mode behaves as if it has mass, 
and the energy is positive even though the stable state is the ground state. 

Once spontaneous symmetry breaking occurs, the ground state has a degree of freedom to move 
through the valley of the Mexican-hat potential.
This means that a new oscillation mode that did not exist before the symmetry was broken appears; it is called the Nambu-Goldstone mode.
The Nambu-Goldstone mode is also called the Nambu-Goldstone boson 
because it is categorized as having boson-like properties as described in the next section.

To mathematically incorporate the concept of spontaneous symmetry breaking into the oscillation model for user dynamics in OSNs, the two following issues need to be resolved.
\begin{enumerate}[$\bullet$]
\item It is very artificial to introduce the Mexican-hat type potential function {\em a priori}, 
so it is necessary to clarify how the Mexican-hat type potential function is inherent to OSNs.
\item It is necessary to clarify what kind of phenomena in user dynamics in OSNs corresponds to the Nambu-Goldstone mode, 
which newly appears due to spontaneous symmetry breaking.
\end{enumerate}

\section{Quantum Theoretic Interpretation of User Dynamics}
In quantum theory, linear operators appear in the theory with two different algebraic rules: 
the commutation relation or the anticommutation relation.
Its feature is that neither is a linear operator commutative in product order.

First, an example of a model that uses the commutation relation is described.
For two linear operators $\hat{a}$ and $\hat{b}$, 
the commutation relation is defined as $[\hat{a}, \, \hat{b}] := \hat{a} \, \hat {b} - \hat{b} \, \hat{a}$. 
In general, $[\hat{a}, \, \hat {b}] \not = 0$.
As an example of such a relationship, if we consider the differential operator $\hat{a} := \dd/\dd t$ with respect to time and time $\hat{b} := t $, their commutation relation is expressed as 
 \[
\left[\frac{\dd}{\dd\,t},\,t\right] := \frac{\dd\,t}{\dd\,t} + t\,\left(\frac{\dd}{\dd\,t}\right) - t\,\left(\frac{\dd}{\dd\,t}\right) = \frac{\dd\,t}{\dd\,t} = 1.
 \]
Since we are considering linear operators that are not ordinary numbers, 
it is not surprising that in general, the products are not commutative. 
We can recognize that this is compatible with matrix representations.
The fundamental equation~(\ref{eq:fundamental-1}) (that is, the expression~(\ref{eq:fundamental-0})) 
is a model of this type.
Such an oscillation mode modeled using the algebraic rules of the commutation relation is called a boson\footnote{This is also called the Bose particle and follows Bose-Einstein statistics.}.

Next, an example of a model that uses the anticommutation relation is described.
The anticommutation relation is defined as $\{\hat{a},\,\hat{b}\} := \hat{a}\,\hat{b} + \hat{b}\,\hat{a}$. 
A model using the anticommutation relation is described by a linear operator satisfying the following relationships: 
\[
\{\hat{a},\,\hat{a}\} = 0, \quad \{\hat{b},\,\hat{b}\} = 0,\quad  \{\hat{a},\,\hat{b}\} = 1. 
\]
From the above relationships, nilpotent characteristics $\hat{a}^2 = 0$ and $\hat{b}^2 = 0$ are obtained.
Such an oscillation mode modeled using the algebraic rules of the anticommutation relation is called a fermion\footnote{This is also called the Fermi particle and follows Fermi-Dirac statistics.}.
One of the most remarkable characteristics of fermions is that no two or more fermions can assume the same quantum state; this is known as the Pauli exclusion principle.
Generally, models described by anticommutation relations have been considered to be unique to quantum theory.

Here, let us look at the anticommutation structure hidden in the fundamental equation~(\ref{fundamenrtal-eq_std}) of OSNs.
Using the definition of $\bm{\mathcal{H}}$ of (\ref{eq:def_H}), $\bm{\mathcal{\Hat{H}}}$ of (\ref{eq:hat_H}) is transformed as follows.
\begin{align}
\bm{\mathcal{\Hat{H}}} &=  \sqrt{\bm{\mathcal{D}}} \otimes
\begin{bmatrix}
+1 & 0\\
0 & -1
\end{bmatrix}
+(\bm{\mathcal{H}} - \sqrt{\bm{\mathcal{D}}}) \otimes \frac{1}{2}
\begin{bmatrix}
+1 & +1\\
-1 & -1
\end{bmatrix}
\notag\\
&=  \bm{\mathcal{H}} \otimes \frac{1}{2}
\begin{bmatrix}
+1 & +1\\
-1 & -1
\end{bmatrix}
+\sqrt{\bm{\mathcal{D}}} \otimes \frac{1}{2}
\begin{bmatrix}
+1 & -1\\
+1 & -1
\end{bmatrix}. 
\end{align}
Using the above expression, the fundamental equation can be written as 
\begin{align}
\ii \, \frac{\dd}{\dd t} \, \bm{\Hat{x}}(t) \! &= \!\! \left(\bm{\mathcal{H}} \otimes \frac{1}{2} \,
\begin{bmatrix}
+1 & +1\\
-1 & -1
\end{bmatrix}
+ \sqrt{\bm{\mathcal{D}}} \otimes \frac{1}{2}
\begin{bmatrix}
+1 & -1\\
+1 & -1
\end{bmatrix}
\right) \bm{\Hat{x}}(t). 
\label{eq:fundamental-2}
\end{align}
Here, by defining  
\[
\bm{\hat{a}} := \frac{1}{2} \,
\begin{bmatrix}
+1 & +1\\
-1 & -1
\end{bmatrix}, \quad 
\bm{\hat{b}} := \frac{1}{2} \,
\begin{bmatrix}
+1 & -1\\
+1 & -1
\end{bmatrix}, \quad 
\bm{\hat{e}} :=
\begin{bmatrix}
+1 & 0\\
0 & +1
\end{bmatrix},
\]
we find the following anticommutation relations: 
\begin{align}
\{\bm{\hat{a}},\bm{\hat{b}}\} := \bm{\hat{a}}\bm{\hat{b}} + \bm{\hat{b}}\bm{\hat{a}} = \bm{\hat{e}}, \quad \bm{\hat{a}}^2 = \bm{\hat{b}}^2 = \bm{\mathrm{O}}\,\text{(null matrix)}. 
\end{align}
From this result, it can be seen that the user dynamics described by the fundamental equation~(\ref{eq:fundamental-2}) (that is, the expression (\ref{fundamenrtal-eq_std})) can be interpreted as exhibiting fermion-like behavior.
Note that this result is derived not from an artificial application of quantum theory.
While pursuing an explicit description of the causal relationship between the OSN structure and user dynamics,  the presence or absence of links represented by the Hamiltonian appearing in the fundamental equation is required to be maintained in the OSN structure, and as a result, fermion-like characteristics are naturally obtained.  
The Pauli exclusion principle is interpreted as stating there is no extra link from $\bm{\mathcal{\Hat{H}}}^2$ other than the link structure of the OSN structure of $\bm{\mathcal{L}}$. 

\section{Solutions of the Fundamental Equations of User Dynamics}
According to the literature~\cite{Ikeya}, we show the closed-form solution of the fundamental equation~(\ref{eq:fundamental-2}).
Also, we show the closed-form solution of the fundamental equation~(\ref{eq:fundamental-1}).

\begin{figure*}
\begin{center}
\includegraphics[width=0.8\linewidth]{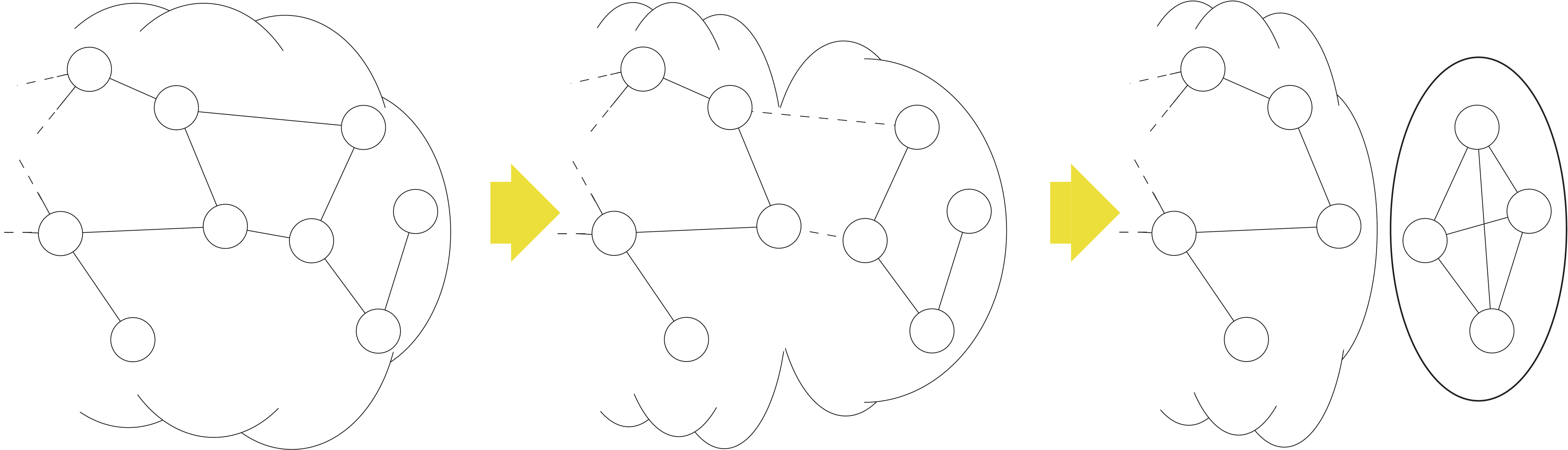}
\caption{Structural changes in OSN due to polarization (1): Formation of a small cluster}
\label{fig:4}
\end{center}
\end{figure*}
\begin{figure*}
\begin{center}
\includegraphics[width=0.8\linewidth]{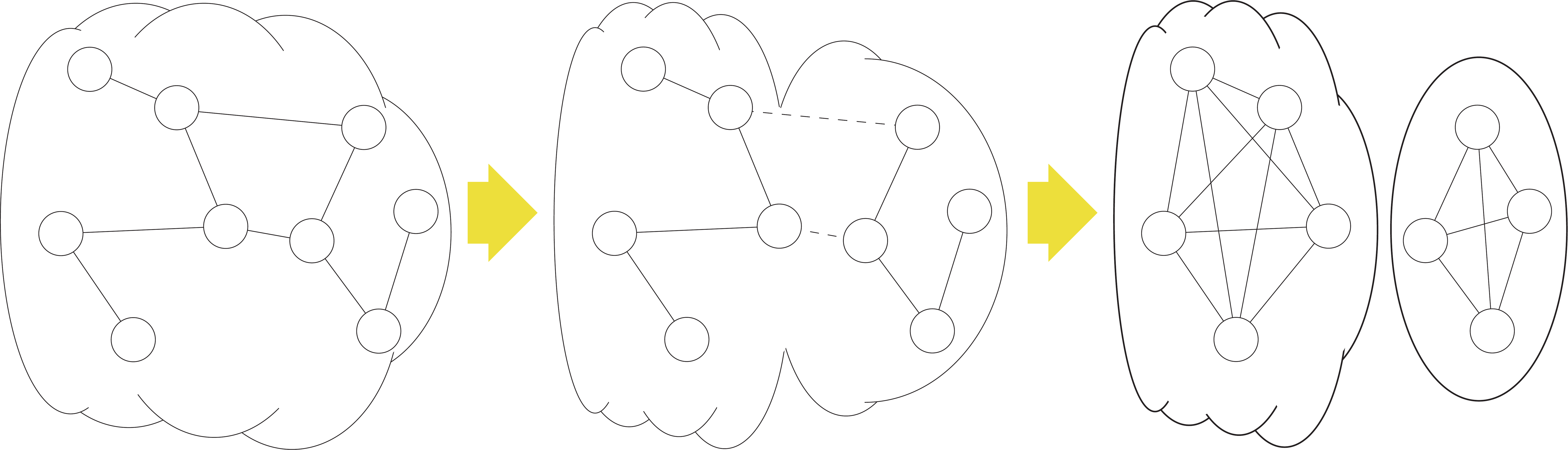}
\caption{Structural changes in OSN due to polarization (2): Polarization yielding two clusters}
\label{fig:5}
\end{center}
\end{figure*}

The solution of the fundamental equation~(\ref{eq:fundamental-2}), is formally expressed as
\begin{align}
\bm{\Hat{x}}(t) &= \exp[-\ii\bm{\mathcal{\Hat{H}}}\,t] \, \bm{\Hat{x}}(0)
\notag\\
&= \left(\bm{\Hat{I}}  - \ii\bm{\mathcal{\Hat{H}}}\,t -\frac{1}{2!}\,\bm{\mathcal{\Hat{H}}}^2\,t^2 + \ii\frac{1}{3!}\,\bm{\mathcal{\Hat{H}}}^3\,t^3 + \cdots \right) \, \bm{\Hat{x}}(0), 
\end{align}
where $\bm{\Hat{I}} := \bm{I}\otimes\bm{\hat{e}}$ is the $2n\times 2n$ unit matrix. 
The expansion of the exponential function on the right-hand side can be expressed relatively easily 
using $\bm{\mathcal{\Hat{H}}} = \bm{\mathcal{H}} \otimes \bm{\hat{a}} + \sqrt{\bm{\mathcal{D}}} \otimes \bm{\hat{b}}$ 
and the anticommutation relation of $\bm{\hat{a}}$ and $\bm{\hat{b}}$.
From the anticommutation relation of $\bm{\hat{a}}$ and $\bm{\hat{b}}$, we have 
\begin{align}
\bm{\hat{a}} \bm{\hat{b}} \bm{\hat{a}} &=(\bm{\hat{e}} - \bm{\hat{b}} \bm{\hat{a}})\, \bm{\hat{a}} =\bm{\hat{a}} - \bm{\hat{b}} \bm{\hat{a}}^2 = \bm{\hat{a}}, 
\notag\\
\bm{\hat{b}} \bm{\hat{a}} \bm{\hat{b}} &=(\bm{\hat{e}} - \bm{\hat{a}} \bm{\hat{b}}) \, \bm{\hat{b}} = \bm{\hat{b}} - \bm{\hat{a}} \bm{\hat{b}}^2 = \bm{\hat{b}}, 
\notag\\
\bm{\hat{a}} \bm{\hat{b}} \bm{\hat{a}} \bm{\hat{b}} &=(\bm{\hat{e}} -\bm{\hat{b}} \bm{\hat{a}}) \, \bm{\hat{a}} \bm{\hat{b}} = \bm{\hat{a}} \bm{\hat{b}}, 
\notag\\
\bm{\hat{b}} \bm{\hat{a}} \bm{\hat{b}} \bm{\hat{a}} &=(\bm{\hat{e}} - \bm{\hat{a}} \bm{\hat{b}}) \, \bm{\hat{b}} \bm{\hat{a}} = \bm{\hat{b}} \bm{\hat{a}} 
\notag
\end{align}
Therefore, when $\bm{\mathcal{\Hat{H}}}^n$ is expanded, the matrices that appear to the right of 
the Kronecker product are just 
\[
\bm{\hat{a}} \bm{\hat{b}} =
\frac{1}{2} \,
\begin{bmatrix}
+1 & -1\\
-1 & +1
\end{bmatrix}, \quad \text{and}\quad
\bm{\hat{b}} \bm{\hat{a}} =
\frac{1}{2} \,
\begin{bmatrix}
+1 & +1\\
+1 & +1
\end{bmatrix},
\]
in addition to $\bm{\hat{a}}$, $\bm{\hat{b}}$, and $\bm{\hat{e}}$.
By using $\bm{\mathcal{L}} = \sqrt{\bm{\mathcal{D}}} \, \bm{\mathcal{H}}$, 
the expansion of $\bm{\mathcal{\Hat{H}}}^n$ is expressed as 
\begin{align}
\bm{\mathcal{\Hat{H}}}^2 &= (\bm{\mathcal{H}} \otimes \bm{\hat{a}} + \sqrt{\bm{\mathcal{D}}} \otimes \bm{\hat{b}})\,( \bm{\mathcal{H}} \otimes \bm{\hat{a}} + \sqrt{\bm{\mathcal{D}}} \otimes \bm{\hat{b}})
\notag\\
&= \bm{\mathcal{H}} \, \sqrt{\bm{\mathcal{D}}} \otimes \bm{\hat{a}} \bm{\hat{b}}+ \bm{\mathcal{L}} \otimes \bm{\hat{b}} \bm{\hat{a}}, 
\notag\\
\bm{\mathcal{\Hat{H}}}^3 &= ( \bm{\mathcal{H}} \, \sqrt{\bm{\mathcal{D}}} \otimes \bm{\hat{a}} \bm{\hat{b}}+  \bm{\mathcal{L}} \otimes \bm{\hat{b}} \bm{\hat{a}})\,( \bm{\mathcal{H}} \otimes \bm{\hat{a}} +  \sqrt{\bm{\mathcal{D}}} \otimes\bm{\hat{b}})
\notag\\
&= \bm{\mathcal{H}} \,\bm{\mathcal{L}} \otimes \bm{\hat{a}} +  \bm{\mathcal{L}} \,\sqrt{\bm{\mathcal{D}}} \otimes \bm{\hat{b}},
\notag\\
\bm{\mathcal{\Hat{H}}}^4 &= (\bm{\mathcal{H}} \,\bm{\mathcal{L}} \otimes \bm{\hat{a}} +  \bm{\mathcal{L}} \, \sqrt{\bm{\mathcal{D}}} \otimes \bm{\hat{b}})\,( \bm{\mathcal{H}} \otimes \bm{\hat{a}} +  \sqrt{\bm{\mathcal{D}}} \otimes \bm{\hat{b}})
\notag\\
&=  \bm{\mathcal{H}} \, \bm{\mathcal{L}} \, \sqrt{\bm{\mathcal{D}}} \otimes \bm{\hat{a}}\bm{\hat{b}}+ \bm{\mathcal{L}}^2 \otimes \bm{\hat{b}}\bm{\hat{a}},
\notag\\
\bm{\mathcal{\Hat{H}}}^5 &= ( \bm{\mathcal{H}} \, \bm{\mathcal{L}} \, \sqrt{\bm{\mathcal{D}}} \otimes \bm{\hat{a}} \bm{\hat{b}} + \bm{\mathcal{L}}^2 \otimes \bm{\hat{b}}\bm{\hat{a}})\,( \bm{\mathcal{H}} \otimes \bm{\hat{a}} +  \sqrt{\bm{\mathcal{D}}} \otimes \bm{\hat{b}})
\notag\\
&= \bm{\mathcal{H}} \, \bm{\mathcal{L}}^2 \otimes \bm{\hat{a}} +   \bm{\mathcal{L}}^2 \, \sqrt{\bm{\mathcal{D}}} \otimes \bm{\hat{b}},
\notag\\
\bm{\mathcal{\Hat{H}}}^6 &= \bm{\mathcal{H}} \, \bm{\mathcal{L}}^2 \, \sqrt{\bm{\mathcal{D}}} \otimes \bm{\hat{a}} \bm{\hat{b}} + \bm{\mathcal{L}}^3 \otimes \bm{\hat{b}} \bm{\hat{a}},
\notag\\
\bm{\mathcal{\Hat{H}}}^7 &= \bm{\mathcal{H}} \, \bm{\mathcal{L}}^3 \otimes \bm{\hat{a}}+   \bm{\mathcal{L}}^3 \, \sqrt{\bm{\mathcal{D}}} \otimes \bm{\hat{b}}. 
\notag
\end{align}
In general, by using $\bm{\mathcal{H}} = \sqrt{\bm{\mathcal{D}}^{-1}} \, \bm{\mathcal{L}}$ we have 
\begin{align}
\bm{\mathcal{\Hat{H}}}^{2k} &= \sqrt{\bm{\mathcal{D}}^{-1}} \, \bm{\mathcal{L}}^k \, \sqrt{\bm{\mathcal{D}}}\otimes \bm{\hat{a}}\bm{\hat{b}}+ \bm{\mathcal{L}}^k \otimes \bm{\hat{b}} \bm{\hat{a}},
\notag\\
\bm{\mathcal{\Hat{H}}}^{2k+1} &= \bm{\mathcal{H}} \, \bm{\mathcal{L}}^k \otimes \bm{\hat{a}} +   \bm{\mathcal{L}}^k \, \sqrt{\bm{\mathcal{D}}} \otimes \bm{\hat{b}},
\notag
\end{align}
for $k \ge 0$. 
Using these relations, the closed-form solution of the fundamental equation~(\ref{eq:fundamental-2}) is given by
\begin{align}
\bm{\Hat{x}}(t) &= \exp[-\ii\bm{\mathcal{\Hat{H}}}\,t] \, \bm{\Hat{x}}(0) 
\notag\\
&= (\cos(\bm{\mathcal{\Hat{H}}}\,t) -\ii\,\sin(\bm{\mathcal{\Hat{H}}}\,t))\, \bm{\Hat{x}}(0)
\notag\\
&= \Big(\sqrt{\bm{\mathcal{D}}^{-1}} \, \bm{P} \, \cos(\bm{\Omega}\,t) \, \bm{P}^{-1} \, \sqrt{\bm{\mathcal{D}}} \otimes \bm{\hat{a}}\bm{\hat{b}}
\notag\\
&\qquad + \bm{P} \, \cos(\bm{\Omega}\,t) \, \bm{P}^{-1} \otimes \bm{\hat{b}}\bm{\hat{a}}\Big) \, \bm{\Hat{x}}(0) 
\notag\\
&\quad -\ii \, \Big(\sqrt{\bm{\mathcal{D}}^{-1}} \, \bm{P} \, \bm{\Omega}\,\sin(\bm{\Omega}\,t) \, \bm{P}^{-1} \otimes \bm{\hat{a}}
\notag\\
&\qquad + \bm{P} \,  \bm{\mho} \, \sin(\bm{\Omega}\,t) \, \bm{P}^{-1} \, \sqrt{\bm{\mathcal{D}}} \otimes \bm{\hat{b}} \Big) \, \bm{\Hat{x}}(0) ,
\label{eq:sol-F}
\end{align}
where $\bm{P}$ is a regular matrix consisting of the eigenvectors of the Laplacian matrix $\bm{\mathcal{L}}$. 
For eigenvalues $\lambda_\mu$ $(\mu=0,\,1,\,\dots,\,n-1)$ of $\bm{\mathcal{L}}$ and the diagonal matrix defined as 
\[
\bm{\Lambda} := \text{diag}(0,\,\lambda_1,\,\dots,\,\lambda_{n-1}),
\]
where $\lambda_0 = 0$. 
$\bm{\mathcal{L}}$ is diagonalized as  
\[
\bm{\Lambda} =  \bm{P}^{-1} \,  \bm{\mathcal{L}} \, \bm{P}. 
\]
In addition, 
\[
\bm{\Omega} := \sqrt{\bm{\Lambda}} = \text{diag}(\omega_0,\,\omega_1,\,\dots,\,\omega_{n-1}),
\]
and, for eigenvalue $\lambda_\mu$ of $\bm{\mathcal{L}}$, $\omega_\mu = \sqrt{\lambda_\mu}$. 
Moreover, 
\[
\bm{\mho} := \text{diag}(0,\,1/\omega_1,\,\dots,\,1/\omega_{n-1}). 
\] 
(\ref{eq:sol-F}) is a solution corresponding to fermions. 

On the other hand, a solution of the fundamental equation~(\ref{eq:fundamental-0}) is given by 
\begin{align}
\label{eq:NG-mode}
\bm{x}^\pm(t) &= \bm{P} \, \exp(\mp\ii\,\bm{\Omega}\,t) \, \bm{P}^{-1} \, \bm{x}^\pm(0), 
\\
&\qquad\qquad \text{(double-sign corresponds)}. \notag
\end{align}
Rewriting the above solution as the solution of the fundamental equation~(\ref{eq:fundamental-1}) gives the following:
\begin{align}
\bm{\hat{x}}(t) &= \Big(\bm{P} \, \exp(-\ii\,\bm{\Omega}\,t) \, \bm{P}^{-1} \otimes 
\begin{bmatrix}
+1 & 0\\
0 & 0
\end{bmatrix}
\notag\\
&\quad\quad\quad + \bm{P} \, \exp(+\ii\,\bm{\Omega}\,t) \, \bm{P}^{-1} \otimes 
\begin{bmatrix}
0 & 0\\
0 & +1
\end{bmatrix}
\Big)\, \bm{\Hat{x}}(0) 
\label{eq:NG-mode2}
\end{align}
This is a solution corresponding to bosons. 

\section{Polarization Model and the Nambu-Goldstone Mode}
Let us assume the polarization of OSNs is caused by spontaneous symmetry breaking. 
To establish this as an acceptable theory, instead of introducing the Mexican-hat potential function shown in Fig.~\ref{fig:SSB}, 
{\em a priori}, we consider a framework of the structural changes of OSNs that inevitably leads to polarization. 
In particular, if the polarization is based on spontaneous symmetry breaking, new dynamics corresponding to 
the Nambu-Goldstone mode, which had not existed before symmetry breaking, should emerge with the polarization.
Therefore, we consider which oscillation mode the appearance of the Nambu-Goldstone mode corresponds to, 
and what kind of phenomenon it corresponds to in OSNs. 

Polarization is a process in which users who have some opinions on a particular topic (for which pros and cons are equally plausible) form separate groups, weakening the interaction between users belonging to different groups, and at the same time strengthening the relationship among users within the same group. 
The presence or absence of a link between users corresponds to the presence or absence of interaction between users. 
This is not limited to direct communication between two users. 
For example, if some information is exchanged on a common electronic bulletin board, it can be considered that some interaction occurs among users that access the board without a direct connection. 
Therefore, it is not unrealistic to think that engagement with an electronic bulletin board, which generally focuses on one common topic, creates links between board participants that yield a complete graph for the said participants.
Therefore, we assume that the following two changes occur simultaneously in the OSN structure due to polarization. 
\begin{enumerate}[$\bullet$]
\item OSN is divided into subnetworks of relatively small size. 
\item The structure of each divided subnetwork is changed and form a complete graph.
\end{enumerate}
Figures~\ref{fig:4} and \ref{fig:5} schematically show this situation.
These indicate situations in which the links indicated by the dashed lines, which express relatively weak relationships, 
are disconnected, and the divided subnetwork becomes a complete graph.  
Figure~\ref{fig:4} shows the situation that the only small subgraph changes to a complete graph, 
and Fig.~\ref{fig:5} shows the situation that both subgraphs change to complete graphs. 

In a normal OSN without polarization, the link structure is sparse, and it is unlikely that the link structure will become a complete graph.
In other words, it is unlikely that all users throughout the world have the relationship of acquaintances or in a relationship directly affecting each other through SNS.
Therefore, the solution~(\ref{eq:sol-F}) of the fundamental equation~(\ref{eq:fundamental-2}) is suggested to model OSN dynamics, 
and the solution~(\ref{eq:NG-mode2}) of the fundamental equation~(\ref{eq:fundamental-0}) cannot exist.

\begin{figure}[tb]
\begin{center}
\includegraphics[width=0.98\linewidth]{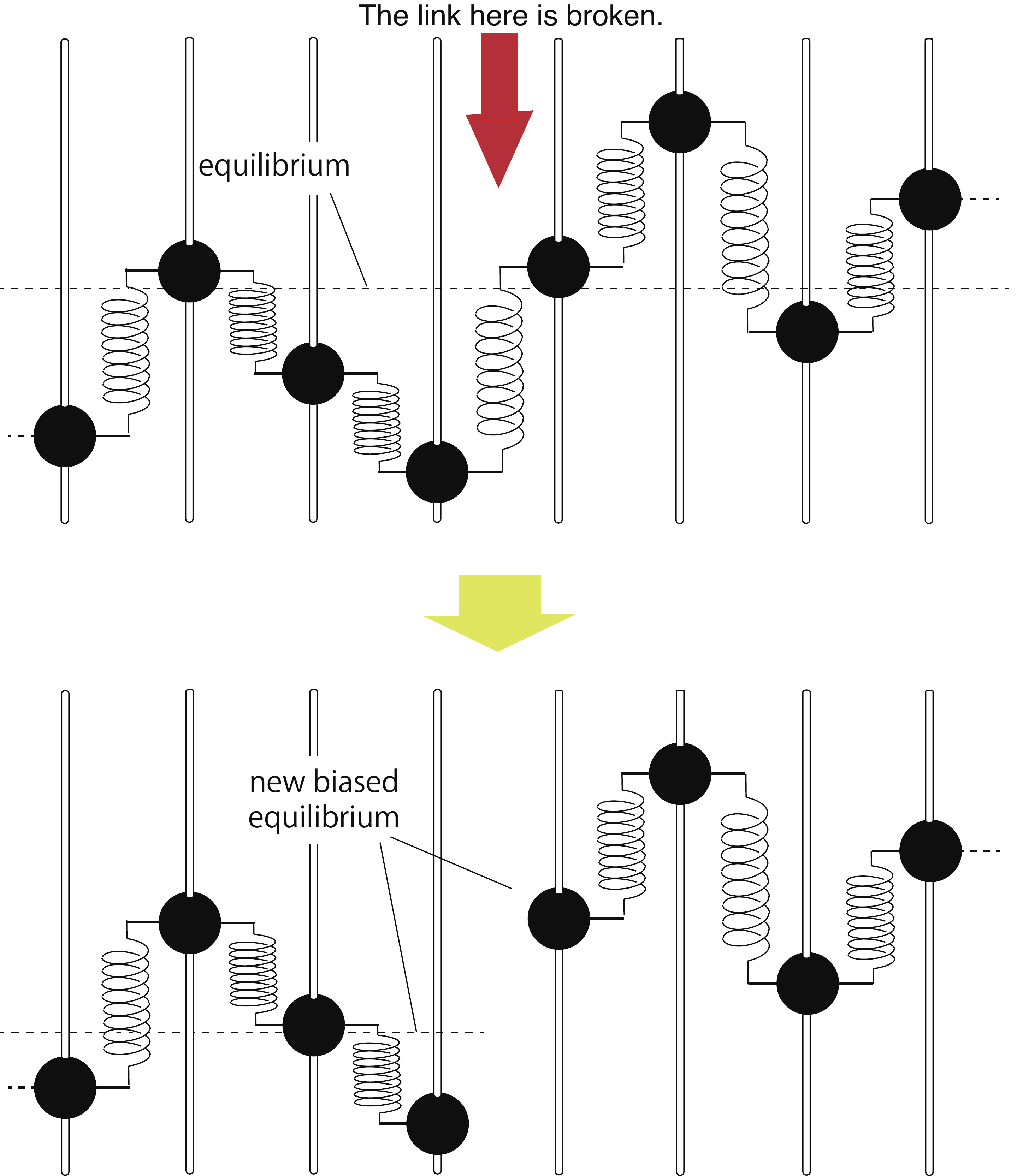}
\caption{Emerging biased equilibrium points caused by fragmentation of OSN}
\label{fig:6}
\end{center}
\end{figure}

Now suppose that polarization breaks the links and causes the emergence of subnetworks.
For certain special cases, we can depict the essence of the oscillation model mechanically by using springs. 
For the special case of a one-dimensional network, let us consider the division into subnetworks 
using a spring expression (Fig.~\ref{fig:6}).
When the whole OSN is connected, the origin of the coordinates denoted by the broken line in the above panel of Fig.~\ref{fig:5} is the equilibrium point. 
If this is split into two subnetworks due to the disconnection of a link, each will have a new equilibrium point. 
This phenomenon means that the stable state is changed from its original position before the split 
to points distant from the original position after the split. 
This emergence of a new biased equilibrium point can be interpreted as a consequence of spontaneous symmetry breaking: 
a new ground state is selected by the Mexican-hat potential shown in Fig.~\ref{fig:SSB}.
This means that the dynamics have positive energy even if the intensity of oscillation is $0$, 
which means that the oscillation mode has mass, 
and it can be interpreted that the activity of the user dynamics increases with polarization.

Furthermore, since the divided subnetwork becomes a complete graph, the solution~(\ref{eq:NG-mode2}) of the fundamental equation~(\ref{eq:fundamental-0}), which did not exist in sparse OSNs, can exist.
This solution constitutes a user dynamics that is created by polarization and is an oscillational solution that can move in a circular motion on the complex plane, so it can be interpreted as a new oscillation mode orbiting the valley of the Mexican-hat potential.
Therefore, it is considered that this newly generated oscillation mode is the Nambu-Goldstone mode that occurs due to spontaneous symmetry breaking.

Finally, we explain the interpretation of the Nambu-Goldstone mode that appears in the complete subnetwork.
Since the Nambu-Goldstone mode does not exist in any sparse OSN, it is not based on the link-based relationships among users, 
but rather depends on the entirety of the subnetwork acting as a coherent entity.  
Therefore, we can recognize that the Nambu-Goldstone mode corresponds to the {\em atmosphere of the field} 
having influence over a closed community. 

\section{Conclusions}
\label{sec:conclusion}
This paper discusses an engineering model for understanding the social polarization in OSNs based on the structure of OSNs; the innovation lies in applying the concept of spontaneous symmetry breaking to the oscillation model of user dynamics in OSNs.
Social polarization in OSNs is defined as splitting the community into relatively small subnetworks and the structure of a split subnetwork becoming a complete graph. 
This allows us to associate the polarization created by spontaneous symmetry breaking with the Mexican-hat potential. 
This work also shows how the Nambu-Goldstone mode is present in the framework of the oscillation model.
To investigate the user dynamics caused by polarization, we should examine the interaction between the fermionic oscillation mode, which always exists in both sparse and complete graphs of OSNs, and the Nambu-Goldstone mode, which is a bosonic oscillation mode existing only in complete OSN graphs. 

\section*{Acknowledgments}
This research was supported by Grant-in-Aid for Scientific Research (B) No.~19H04096 (2019--2021) and No.~20H04179 (2020--2022), 
and Grant-in-Aid for Scientific Research (C) No.~18K11271 (2018--2020)  from the Japan Society for the Promotion of Science (JSPS).



\end{document}